%% file: main.tex
\documentclass{acm}

\usepackage{times}  
\usepackage{array, graphics, graphicx, color, url, xspace, multirow, rotating, epsfig, amsmath, amsfonts}
\usepackage{wrapfig}
\usepackage{tikz}
\usepackage{epstopdf,subfig}
\usepackage[format=plain,labelfont=bf,font=small]{caption}
\usepackage{ragged2e,flushend,enumitem,listings}
\usepackage{blindtext}
\usepackage{acronym} 
\usepackage{hyperref, cleveref}
\usepackage{multirow}
\usepackage{svg}
\usepackage{appendix}
\usepackage{tabularx}
\usepackage[small,compact]{titlesec}

\let\svthefootnote\thefootnote
\newcommand\freefootnote[1]{%
  \let\thefootnote\relax%
  \footnotetext{#1}%
  \let\thefootnote\svthefootnote%
}

\usepackage{listings}[mathescape]
\usepackage{color}
\usepackage{outlines}
\usepackage{pifont}

\hypersetup{pdfstartview=FitH,pdfpagelayout=SinglePage}

\newcommand\paragraphb[1]{\noindent{\bf{#1}}}
\newcommand\paragraphi[1]{\noindent\emph{#1}}
\newcommand\pb[1]{\paragraphb{#1}}
\renewcommand\pi[1]{\paragraphi{#1}}

\newcommand{\bi}{\begin{itemize}}
\newcommand{\ei}{\end{itemize}}

\newcommand{\eat}[1]{}

\newcommand{\allnotes}[1]{}
\renewcommand{\allnotes}[1]{\textit{#1}}
\newcommand{\figspace}{\vspace{-10pt}}

\graphicspath{{./dia/}}
\begin{document}

\title{DBNet: Leveraging DBMS for Network Automation}
\author{Rithvik Chuppala and Silvery Fu (co-first authors), Sylvia Ratnasamy \\ UC Berkeley}
\maketitle

\begin{abstract}
We present DBNet, a data-driven network automation framework built on top of a DBMS. DBNet utilizes key primitives of a DBMS including tables, procedures, transactions, logging, and access control to serve the functions of a data-centric network control plane. DBNet accomplishes this functionality by storing mirrored network device states, executing automation programs on these mirror states within the DBMS, and proxying state updates out to the physical devices upon changes to mirror/local state. The framework also stores network telemetry data, performs analytics on the data, uses the analytics to motivate control plane actions, and provides provenance logging features on the actions taken. We apply DBNet to motivating cloud network infrastructure examples and show how developers can use DBNet’s interface to express rich user-defined policies. Our preliminary case studies show that the overhead to run DBNet is trivial in the timescales generally relevant for network automation.
\end{abstract}

\input{body}

\bibliographystyle{abbrv} 
\bibliography{main}

\end{document}

%% file: body.tex
\section{Introduction}

A growing trend in the network infrastructure and IoT framework spaces is the paradigm of data-driven automation - using data collected from the system of interest to motivate control and management actions taken on the system. A number of reasons have contributed to this trend, including the growing centralization of data collection in the computing field at large \cite{OCIData} in addition to increasing systems support to process this data for downstream usage \cite{Ray} \cite{Spark} \cite{MapReduce}. This trend represents the gradual yet continued evolution of the network control plane. What began as a static operator-controlled interface with tightly coupled control and data plane policies evolved into a multi-layer abstraction-based SDN interface providing networks with a degree of runtime automation \cite{Road}. The next step in this evolution is a data-centric approach that combines in-network data sources with the aforementioned runtime platform \cite{Flex}. By utilizing real-time data from the network, the automation policies expressed through the runtime platform can target actual network conditions, improving efficacy of control operations. Instead of a trial-and-error approach where operators cycle through different policies and find the one that works best empirically \cite{NetAutoVideo}, the data-driven approach organically suggests this policy by using data analytics and the insight derived from analytics. Automation policies based on network data can detect and react to network failures, device anomalies, and policy violations. We observe the trend of data-driven automation in the space of IoT frameworks as well \cite{Dspace}. 

The process of data-driven automation can be broadly broken down into four major components: observability, analytics, automation, and provenance. This decomposition represents a very common theme in IoT and other data-centric tasks: first a sensory functionality examines the behavior and state of our system; and then an actuation functionality makes control decisions based on the observations \cite{DDN}. 

The first step of any data-driven process is collecting the data itself - observability is the notion of having visibility into the network to collect this data. Observability allows users to collect data on the state and behavior of the network. This state can include network device state, network topology information, network routing state, and other key measures such as link utilization. As cloud continues to blur the distinction between compute infrastructure and network infrastructure, figures such as CPU utilization, memory usage, and ingress traffic load are also important for network observability. A key aspect of observability is telemetry, which refers to the medium through which data on network state and behavior is presented to the user. Telemetry commonly comes in the form of traces (a manifestation of paths taken through the network), logs, and metrics \cite{Timescale}. 

The next step of data-driven automation is to perform analytics on the network data gathered through telemetry \cite{Timescale}. The objective of analytics is to transform the collected network data into something that provides the user or operator with proper insight into network behavior. This requires systems to store all the data collected through telemetry, platforms to query the data and extract the necessary metrics, and systems to execute this query in an efficient manner. In the status quo, telemetry data is generally streamed to a dedicated storage system that can either serve queries on the data itself or send it to yet another dedicated analytics system that serves queries \cite{Snicket} \cite{Sonata}. 

The third step is automation. Automation uses data insights to inform and motivate network actions. Without data, automation would simply involve programming network devices and components. However, by leveraging data and analytics, automation involves calculating the relevant metrics for the task and then performing actions based on whether certain conditions are met in the network~\cite{Flex}. Not only does data-driven automation motivate actions taken on the network, but it also increases the explainability of actions since modified network state can be attributed to prior network behaviors and conditions. This is important for the fourth step.

The final component of data-driven automation is action provenance. Data provenance helps us answer the questions of ``what happened'', ``why it happened'', and “who made it happen.'' \cite{InternetScaleProvenance} To answer these questions it is necessary to recursively explain a particular result with its causes, and then explain their own causes, until we arrive to the root states. A well-known method to provide data provenance is through effective logging. By utilizing change-data-capture processes, the system can record important events such as existence of write operations, the operation itself (what the data used to be and what it has been changed to), the timestamp of the operation, and the user who initiated the operation \cite{DBOSProvenance}. Moreover, provenance can track the programs that actually modify network state and log both calls to those programs as well as the user/parent program that performed the call.

\subsection{Today's Challenges}

We note a few key challenges of today’s systems in the pursuit of data-driven network automation. 

One challenge is managing visibility and telemetry data. There is ongoing research and literature in collecting telemetry data from the network \cite{Snicket} \cite{Sonata} \cite{Spidermon} \cite{PacketLevelTelemetry} but storing, managing, and utilizing this data is difficult \cite{Timescale}. Specifically, telemetry data, when collected, comes in many forms such as logs, traces, and metrics, all of which can come in numerous forms and standards. This raw data needs to first be processed into a form that is easily interpretable. Then, end-users of the data, such as network administrators, engineers, and developers need to make sense of this data to perform their tasks. In other words, the telemetry data has to do more than just provide streams of information on network state; it has to provide end-users of that data with the visibility needed to gather insight and take action. This can involve numerous transformations of the data, streaming in and out of various storage, processing, and visualization systems. End-users are tasked with more than just observing telemetry data; they instead have to manage and develop an end-to-end data pipeline surrounding the data as well.

There is also a distinct lack of a flexible interface for user-defined network infrastructure policies. Currently, if a user wants to create policies to automate provisioning and management of their network infrastructure in the cloud, for example, they are locked to a small selection of in-built cloud provider policies. Users do not have access to a rich interface to define their own policies for their network infrastructure. 

Another challenge facing data-driven network automation is the lack of established provenance schemes in the networking field. As a discipline, provenance originated and has been well-studied in the database fields \cite{DBOSProvenance} but is not a database-specific concept. It is starting to be studied by the networking community as well \cite{GoodBadDiff} but provenance, for the most part, remains an unincluded member of the first-class concerns of network application developers and admins. 

The final challenge that we will discuss, and arguably the most important, is bifurcated systems. Currently, each step of data-driven automation is performed in its own separate platform. For example, a telemetry platform collects data from traces, logs, and metrics and streams the results out to a separate data store. This data store needs to configure its schema to match the output data of the telemetry platform. Then, analytics is performed using yet another OLAP-based data system. The users then have to utilize the analytics from its data system to define automation programs on a network automation tool. This involves massive amounts of data transfer across multiple platforms and makes it very difficult for developers, SREs, and operators to manage and coordinate all the platforms.

\section{Design Objective and Rationale}

\subsection{Problem Definition}

The above challenges motivate the problem definition: 

\begin{center}
\textit{Can we embed the functionalities of a network automation and telemetry system inside of a DBMS?}
\end{center}

We will need to evaluate whether today’s DBMS provides enough performance and primitives in order to successfully orchestrate a telemetry platform, analytics engine, automation platform, and provenance feature, all embedded within the same system. In other words, the idea is to design a data-centric network control plane (automation platform) for network infrastructure and systems. 

The DBNet framework essentially mirrors network device states, executes automation programs on these mirror states within the DBMS, and proxies state updates out to the physical devices upon changes to mirror/local state. Automation policies are implemented using a DBMS interface that modifies the mirror or soft states. Device soft states (mirror states) are stored in tables in the DBMS and automation policies modify these tables. Telemetry data streams into the DBNet framework, gets stored in its own tables, and updates the relevant table metrics. Provenance is achieved through logging all operations on relevant state-containing tables as well as logging automation procedure calls. In other words, a DBMS Interface and its primitives are used to perform the control plane operations in a network automation platform.

\subsection{The DBMS}

The DBMS provides many useful primitives that make it well-suited to serve the purpose of a data-driven network automation platform. We will survey the key DBMS primitives that we can use in the data-centric control plane.

\pb{(1) Tables} Despite being the most basic element of a database management system, tables serve multiple purposes in the DBNet framework. The main purpose is to store network device mirror states, hard states for virtual network devices, and auxiliary network information that aids in automation or analytics functionality. Other key uses of tables are to serve as a log for the provenance functionality and to store streamed-in telemetry data from the network.

\pb{(2) Stored Procedures and Triggers} Stored procedures are mixed declarative and imperative-language procedures that are compiled and stored in the DBMS for reuse. In Postgres, these procedures can combine SQL statements with an imperative language like Python, C++, or Java to offer a rich variety of read/write functionality, query semantics, and application logic. Triggers are used alongside stored procedures. Triggers act on a particular database event---such as a row insert or modification to a table---and execute a stored procedure in the event that this defined ``trigger'' occurs.  

Together, stored procedures and triggers serve as the primary means of automation and control plane executions. For example, stored procedures perform planned modifications device mirror state, sync physical device states with mirror states, and perform conditional modifications of state via triggers. Triggers provide change data capture which serves as logging for provenance. On a call to a stored procedure or write modification to a table, triggers invoke a procedure to add a logging entry to a dedicated log table with the specified procedure call or write modification. Moreover, stored procedures and triggers perform real-time analytics on telemetry data as it is streamed in from the network to DBNet. Overall, stored procedures and triggers provide a rich programmable interface in both declarative and imperative paradigms.

\pb{(3) Transactions} Database transactions wrap around consecutive database operations to provide isolation of execution and atomicity features. If a wrapped group of database operations succeeds, the transaction commits and all its changes are made permanent; if not, the transaction rollbacks and its changes are reverted. When used in DBNet, transactions wrap every stored procedure which follows naturally as stored procedures are simply consecutive database statements. Transactions also wrap around consecutive calls to stored procedures. These ensure atomic and isolated automation behavior with rollback features to a consistent network state if an update fails. 

\pb{(4) Query Language} The query language primarily provides us with a medium through which we can execute analytics queries. Embedding these queries in stored procedures lets other procedures use these queries to conditionally perform network changes.

\pb{(5) Access Control} Access control uses configured rules and policies to determine whether a user can modify a database object. This provides the DBNet framework with data governance features and authorization primitives for data security.

\pb{(6) Constraints} Database constraints provide data integrity, allowing DBNet users to define rules for what values certain tables and columns can take.

\subsection{Related Works}

\pb{Network Modeling and Automation} Onix \cite{Onix}, B4 \cite{BFour}, and Orion \cite{Orion} present network modeling and automation from the lens of an SDN controller with data store internals. Some of these controllers adopt database-inspired designs and show promising operational benefits~\cite{davie2017database}. DBNet can be used in-conjuction with these SDN controllers by layering over the inner state-storing database in the SDN controller and providing additional observability, analytics, and provenance benefits.

\pb{Database-Oriented Systems} DBOS \cite{DBOS} and its related paper Apiary \cite{Apiary} utilize a database-oriented approach to rebuild a computer operating system. The key difference between the works of the DBOS authors and the considerations of DBNet is that DBOS stores all system hard state in the database and DBNet can only store device mirror state. Since the operating system runs as a program, all entities can exist with their actual state or hard state in the database tables. However, since the network is a physical entity, the only mirror states or soft states can be stored in the DBNet database tables. This paradigm difference has motivated different architectures; whereas interacting with DBOS involves writing short-lived executions running in AWS lambda-like environments, we have chosen for DBNet to be built as a stateless but centralized server that proxies updates to physical device states and interacts with the underlying DBMS. Additionally, DBOS utilizes a ``polystore'' scheme, copying and transferring state and data across multiple database systems. DBNet’s ideology is to minimize data movement so a single datastore is used instead.

\pb{Network Telemetry} Snicket \cite{Snicket} and Sonata \cite{Sonata} take a look at query-driven distributed tracing and query-driven streaming network telemetry, respectively. Both utilize a bump-in-the-wire approach, where service proxies running alongside microservices (in the case of Snicket) and/or programmable switches (in the case of Sonata) filter traces based on input queries. Both can be used in-conjunction with DBNet, providing DBNet with filtered and anomalous telemetry data to begin with, that can focus control plane executions around anomalies. However, a well-representative analytics functionality may be a trade-off in this situation.

\section{DBNet Architecture}

\begin{figure}
\centering
\includegraphics[scale=0.33]{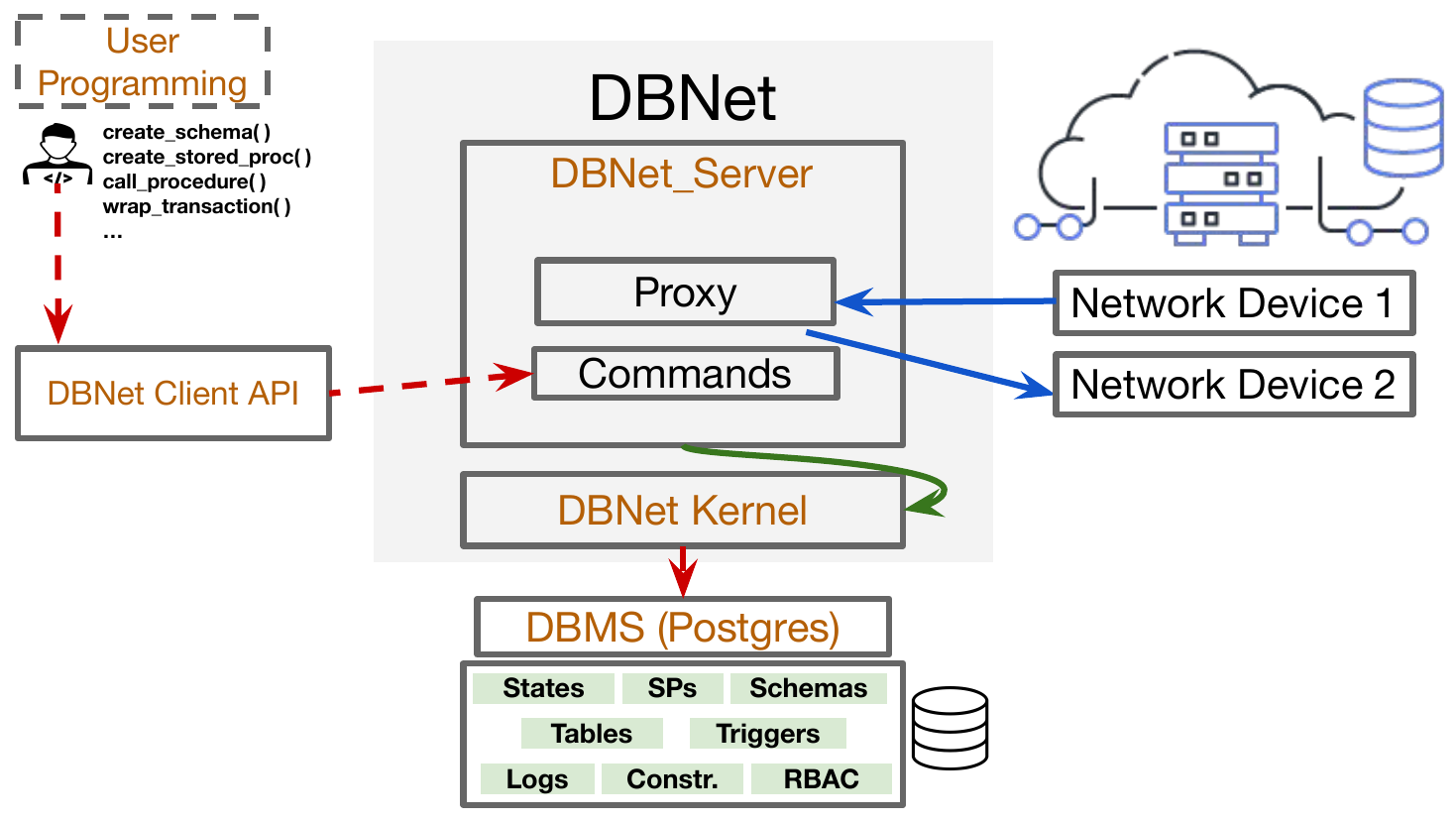}
\caption{DBNet Architecture}
\label{fig:dbot_arch}
\figspace
\end{figure}

DBNet is a stateless system built as a layer on top of the DBMS. As is visible in Figure \ref{fig:dbot_arch}, DBNet itself consists of two layers: the DBNet Server and the DBNet Kernel. The DBNet Server has a twofold purpose. The first is to serve as an HTTP Server with endpoints to accept Client API requests and commands and then to pass on those commands to the DBNet Kernel. The second purpose of the DBNet Server is as a proxy to synchronize physical device states and internal soft states. The DBNet Kernel is a direct interface to the DBMS. It translates client requests from the server to the postgres query language. Finally, the underlying DBMS (Postgres) contains tables, stored procedures, triggers, logs, constraints, and access control. 

\subsection{DBNet Kernel}

The DBNet Kernel serves as the lowest-level interface to the database. All commands that directly interact with SQL to create tables, create procedures, set up schemas are contained within the Kernel. When DBNet starts, the first thing the kernel does is perform “start-up tasks”. These include creating the log table and creating a trigger + procedure to append to the log table upon a Create, Update, Insert, or Delete operation on a table. Other commands in the kernel create schemas, create tables, create stored procedures, create stored procedures with triggers, etc. 

\subsection{DBNet Server}

The DBNet Server is the upper layer of the DBNet stack and serves as a bridge between the external world and the DBNet Kernel. There are three primary ways in which the DBNet Server interacts with the external world: a southbound path from API commands called by client programmers, a southbound path from devices with telemetry data streaming in, and a northbound path to devices to transfer state updates to physical devices. The DBNet Server itself is an HTTP Server with endpoints to accept Client API requests and handle commands. These commands are subsequently passed on to the DBNet Kernel which handles executing them. Before starting the HTTP Server, the DBNet Server also starts up Postgres and establishes a connection to it. The southbound path from devices also utilizes the same HTTP Server and endpoints. The northbound path pushes updates from DBNet to the devices; it does not utilize the HTTP server and instead invokes the proxy.

\subsection{DBNet Client API}

The DBNet Client API is a client API for users to program DBNet. The client API includes commands such as createSchema, createStoredProcedure, callProcedure, wrapTransaction, etc. The client API sends the target information to the DBNet Server which forwards it to the DBNet Kernel that invokes the Postgres commands for intented functionalities. 

\section{Motivating Examples}

\begin{figure}
\includegraphics[scale=0.37]{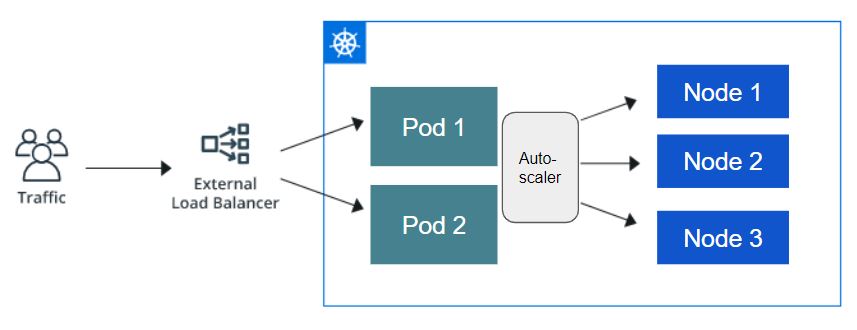}
\caption{Diagram of Nodes, Pods, Auto-scaler, Load-balancer Use Case}
\label{fig:node_pod_diag}
\figspace
\end{figure}

\subsection{Setup}

Consider the scenario where a developer runs an application using a canonical microservice-based architecture. As displayed in the figure, the containerized microservices run in a pod each and all the pods multiplex between several compute nodes available to the cluster. In terms of network infrastructure, an auto-scaler is in charge of provisioning new nodes and killing nodes as policy dictates and a load-balancer that splits ingress traffic amongst the multiple pods and nodes. Additionally, consider that the application is connected to an observability front-end, such as Jaeger or OpenTelemetry. The traces, logs, and metrics from the observability front-end are streamed into DBNet and collected in dedicated tables. 

In DBNet, there exists a table to store state information on each of the pods, nodes, autoscalers, and load balancers. An OOP analogy to describe this approach is that tables generally represent a “Class” and each row represents an “Object” of the table “Class”. As is visible in Table \ref{table:podnodeaslb}, there is a table for Pods and Nodes, a table for Auto-scalers, and a table for Load-balancers; each of these tables are populated with the pods, nodes, auto-scalers, and load-balancers that the current DBNet instance is in charge of controlling. The schema and table that holds traces, logs, and metrics are OpenTelemetry \cite{OpenTelSpec} compatible and based on the OpenTelemetry Spans and Traces specs. The tables comprise the system's data plane.

In order to update the state and behavior of the physical network components, the user programs DBNet to modify the states of these components in the row corresponding to the device of interest in the table for that specific component type. For example, if they want to modify the status of a pod, they find the corresponding PodID in the Pods table and change its status. This changed state propagates to the physical pod in the network which updates its internal state to reflect this change. The user programs stored procedures and triggers to implement automation policies to modify state and behavior of network device components. 

\begin {table}
\begin{tabular}{ |p{1cm}|p{1cm}|p{1cm}|p{1.7cm}|  }
 \hline
 \multicolumn{4}{|c|}{Nodes} \\
 \hline
 nodeId & podId & cpuUtil & ingressTraff. \\
 \hline
 1 & 1 & 0.37 & 221 \\
 2 & 1 & 0.82 & 382 \\
 3 & 2 & 0.14 & 67 \\
 ... & .... & ... & ... \\
 \hline
\end{tabular}

\begin{tabular}{ |p{1cm}|p{1.7cm}|  }
 \hline
 \multicolumn{2}{|c|}{Auto-Scalers} \\
 \hline
 podId & avgCpuUtil \\
 \hline
 1 & 0.595 \\
 2 & 0.14  \\
 ... & .... \\
 \hline
\end{tabular}

\begin{tabular}{ |p{1cm}|p{1.7cm}|  }
 \hline
 \multicolumn{2}{|c|}{Load-balancers} \\
 \hline
 podId & avgIngTraff. \\
 \hline
 1 & 301.5 \\
 2 & 67  \\
 ... & .... \\
 \hline
\end{tabular}

\caption{Nodes, Auto-Scalers, and Load-balancers Tables}
\label{table:podnodeaslb}
\figspace
\figspace
\end{table}

\textbf{Example 1: Traffic Load Balancing}

In the first example situation, the developer wants to coordinate the behaviors of load-balancer and auto-scaler. The policy they aim to implement is as follows: if the ingress traffic into the load-balancer exceeds a certain traffic threshold, have the auto-scaler spin up a new node. Then adjust the load-balancer to balance traffic between the old nodes and the additional new node. In the status quo, the cloud provider gives no flexibility to express this policy. The only integration that the cloud provides is to adjust the load-balancer to balance traffic between a newly autoscaler-provisioned node in an eventually consistent manner. Generally, the cloud provider gives the option to select between a few of policies for the auto-scaler, but these are mostly different heuristics for CPU utilization. 

By using DBNet, the developer derives several key benefits in the example. Primarily, DBNet allows for the flexibility of the control plane to express this behavior via stored procedures and triggers. The trigger acts on an update to the load-balancer’s ingress traffic value; if this value starts to exceed a threshold, the triggered stored procedure will change state in the autoscaler to provision a new node. Within the same transaction, the load-balancer state will have access to the newly created node and balance traffic to that node as well. The second benefit here is that the developer can modify state atomically for both the load-balancer and the auto-scaler via transactions. In addition, statistics on network traffic, CPU utilization, time-of-day queries are readily available and easily accessible since the relevant data is all stored in the same DBMS. Provenance logging provides additional benefits. For example, provenance gives answers to questions like “why was this new node provisioned?” By tracing the logs from the start, the developer sees a new entry was added to the Nodes table, which was done by a stored procedure execution, which was called by a trigger activation, which was a result of the ingress traffic into the pods exceeding a certain threshold. 

\textbf{Example 2: Network Telemetry}

In the second example situation, the developer wants to start automating their network infrastructure using telemetry data. Telemetry service traces follow the path of a user-generated trace from one service to another to help with microservice debugging efforts. In the current situation, the developer wants to determine which nodes have the highest error rates and which nodes have the highest service latencies. The autoscaler should kill off nodes with error rates or latencies exceeding a certain threshold and provision new ones, effectively restarting the deployment. In the status quo, there is no straightforward way to integrate telemetry data with network behavior. Telemetry data and network control plane automation exist on two completely different platforms and are very hard to dynamically integrate with each other.

With DBNet, the developer is able to co-locate collected telemetry data, network states, automation procedures which naturally creates an environment offering rich control plane programmability. As with above, transactions, analytics, and provenance logging offer benefits in this situation too. 

\section{Preliminary Performance Benchmarks}

The DBNet framework pairs with existing automation and telemetry frameworks, offering additional benefits on top of the standard functionality. Next, we evaluate the overheads associated with running DBNet.

To benchmark DBNet, we ran it on a local machine, on top of Postgres running locally as well. In order to run the motivating example cases, we ran minikube locally as well and utilized the Kubernetes API to provision the objects. Since everything ran locally, all performance benchmarks are of the critical system path itself, and not networking delays. 

\textbf{Example 1: Traffic Load Balancing}

The first experiment evaluated the end-to-end latency overhead of a real-world DBNet network automation setting. We implemented all the necessary tables, schemas, stored procedures, and triggers for the above motivating use case.

The average latency to create 7 schemas, provision 7 tables, create 9 stored procedures, and 5 triggers for the entire example was 0.35 seconds. This overhead is small and represents a one-time programming of DBNet by the developer.

To implement the load balancing and autoscaling example above, we populated the Nodes table with several nodes belonging to one of several pods and with synthetic CPU utilization and ingress traffic data. Triggers and Stored Procedures would handle populating the Auto-scalers and Load-balancers tables by using Nodes data to calculate avg CPU utilization and ingress traffic per pod. Next, we modified the synthetic data of a few nodes to have their ingress traffic exceed a made-up threshold defined in the stored procedures given the target policy: "if the ingress traffic into the load-balancer exceeds a certain traffic threshold, have the auto-scaler spin up a new node." In a real-world setting, this exact dataflow would be followed by external devices to interact with the DBNet system. The trigger activated once the ingress traffic for a pod exceeded a certain threshold. The corresponding stored procedure would call the Kubernetes API to create a new node, update the DBNet tables, and re-calculate the ingress traffic and cpu utilization metrics. 

The average latency to execute the task in this autoscaling example, from the time the average ingress traffic exceeded the threshold to when the ingress traffic and CPU utilization metrics were recalculated, was approximately 1.8 seconds, with 0.027 seconds taken by all DBNet executions and the rest from the creation of a new node by the Kubernetes API. Similarly, the average latency to execute "Example Situation 2" from the time the anomalous traces came in to the recalculation of metrics was approximately 1.9 seconds, with 0.045 seconds taken by all DBNet executions and the rest by the Kubernetes API. This demonstrates that the overhead for running DBNet is almost negligible compared to the timescales of network automation, especially when considering cloud resources are provisioned on the order of minutes.

\begin{figure}
\centering
\includegraphics[scale=0.6]{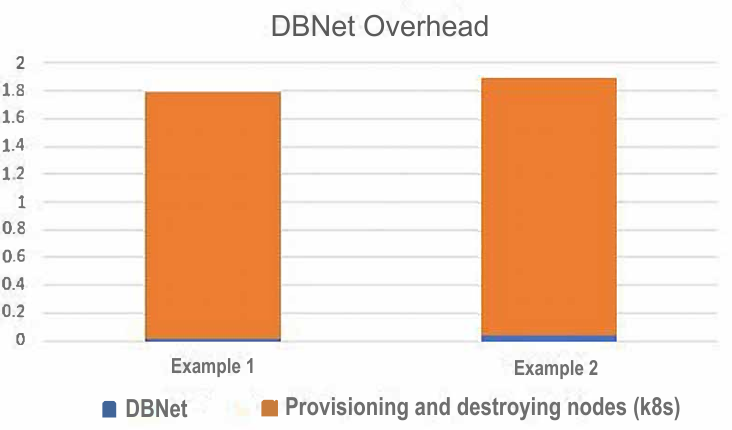}
\caption{Overhead of DBNet in a real-world network automation setting}
\label{fig:dbot_overhead}
\figspace
\end{figure}

\textbf{Case 2: Network Telemetry}

For Example 2, we streamed in synthetic telemetry data after populating the Nodes, Auto-Scalers, and Load-Balancers table. The synthetic telemetry data would exhibit an error rate and latency above the threshold to kick off the creation of a node, table updates, and other stored procedure calls.

For the second experiment, we evaluated the load-handling capabilities of the DBNet system.  In order to test this, we set up 20 clients that each streamed in 100 traces to DBNet at the same time. This represents the main type of load DBNet will handle since this emulates telemetry data streamed in from many sources at the same time. DBNet was able to ingest this telemetry data, store it, and report a success in an average of 3.2 seconds. Since DBNet will likely be run in the cloud, scaled up as necessary, for production use-cases and environments, this represents the lower bound load handling capabilities of DBNet.

\section{Conclusion and Future Directions}

We propose DBNet, a data-centric control plane framework that focuses on network infrastructure automation. We discussed the trends of data-driven network automation and the pain-points faced by developers today. In order to remedy these challenges, we identified key DBMS primitives including tables, stored procedures, triggers, transactions, access control, and constraints. We designed the DBNet architecture focused on remedying the challenges in data-driven networking while also keeping the ideals of our problem definition in mind. We provided a review of related work and have shown how key DBMS primitives can provide a plethora of benefits through motivating use cases and preliminary performance benchmarks. 

Given our proposal and initial findings, there are many important questions require answering for applying DBNet's approach in practical settings, from performance, scalability, to its impacts on existing and future network systems. We discuss these future research directions in what follows.

\pb{Understanding performance.} An important aspect of DBNet is its performance, particularly in the context of varying network scales and complexities. Benchmarking and understanding the latency and overhead of DBNet in different use cases, from small-scale local networks to large-scale cloud infrastructures~\cite{Onix,BFour}, are part of our ongoing research. In particular, the performance impacts of additional features such as analytics and provenance logging should also be evaluated comprehensively in these scenarios.

\pb{Scalability and deployment.} As with performance, while the DBNet offers a flexible platform for network automation, its scalability, particularly in large networks with high volumes of telemetry data and numerous devices, is worth investigating~\cite{Onix,BFour,Timescale}. Meanwhile, questions remain open around about the DBNet's implications on load balancing, autoscaling, and other deployment issues, such as monitoring DBNet's service level objectives (SLOs) via distributed tracing and telemetry. 

\pb{DBNet's role in network analytics and provenance.} DBNet's architecture introduces a convenient means of storing and analyzing network data. Its inherent capabilities to perform real-time analytics and log provenance open up new possibilities in network data analysis and auditing. However, we need to better understand the scope of analytics and provenance tasks~\cite{InternetScaleProvenance,Snicket,Sonata} that can be achieved with DBNet and the potential applications in troubleshooting, network optimization, and security.

\pb{Network governance and security.} As DBNet leverages DBMS's access control for data security, it would be interesting to delve into how this could impact overall network security, the handling of sensitive network information, and the applicability of these DBMS primitives in networking context~\cite{DBOS,Apiary}. The potential use of formal methods to ensure data integrity in DBNet's operations and how they tie into the existing access control mechanisms could be an interesting direction for future exploration.

\pb{Extensibility and user-defined policies.} DBNet's programming interface that combines declarative and imperative specification offers room for customization and extension. How developers could express user-defined policies using DBMS interface and primitives could potentially drive innovative applications, such as synthesizing and optimizing the stored procedures towards known topology and tasks~\cite{mogul2020experiences}. The trade-offs involved in extending DBNet, such as maintaining performance while providing such flexibility, is also an interesting topic of exploration. Finally, it's also worth exploring the integration of DBNet with systems for other network environments such as mobile networks~\cite{luo2021democratizing} and IoT~\cite{sosp21-dspace}.

\pb{Impact on existing network systems.} The flexibility of DBNet allows integration with existing network infrastructure, such as incorporating (rather than replacing) the existing network primitives for state mirroring and network telemetry~\cite{Sonata,Onix,BFour}. Yet, it also poses a question of how the transition to DBNet would impact these tools or systems in terms of their performance, scalability, and security, in ways discussed above.

\pb{Impact on future networking designs.} DBNet's data-driven network automation approach could also potentially inform and influence the design of future network control planes and network infrastructure management systems. In particular, how DBNet can contribute to the aspects of network modeling~\cite{mogul2020experiences}, automation, and telemetry, especially with the advent of technologies like AI/ML~\cite{trummer2022codexdb}, presents an exciting avenue for research and exploration.